\documentclass[aps,prd,twocolumn,groupedaddress]{revtex4-2}
\usepackage{amsmath,amssymb,graphicx}
\usepackage{hyperref}
\hypersetup{hidelinks}
\begin{document}


\title{Gravitating kinks with asymptotically flat metrics}
\author{Ziqi Wang}

\author{Yuan Zhong}
 \email{Corresponding author: zhongy@mail.xjtu.edu.cn}
 
\author{Hui Wang}

\affiliation{MOE Key Laboratory for Nonequilibrium Synthesis and Modulation of Condensed Matter, \\ School of Physics, Xi’an Jiaotong University, Xi’an 710049, China}

\date{\today}

\begin{abstract}
	In this work, we consider a two-dimensional (2D) dilaton gravity model where the dilaton kinetic term $\mathcal{X}$ is modified by an additional derivative coupling term  $\alpha\mathcal{X}^2$. In the case with a canonical scalar matter field, the field equations of this model have a simple first-order formalism, from which exact static kink solutions can be constructed. The novelty of these solutions is that the corresponding metric can be asymptotically flat rather than asymptotically anti de Sitter.  The linear stability and the localization of scalar matter fields are also studied. It was found that the solutions are stable against small linear perturbations, and the localization of scalar matter fields can be realized by introducing scalar-kink interactions.
\end{abstract}

\maketitle

\section{Introduction}
Kink solutions in 1+1-dimensional nonlinear scalar field models are probably the simplest topological soliton solutions. Yet kink and its higher-dimensional extension, domain wall, provide us ideal toy models for understanding complicated issues such as the dynamics of cosmic vacuum bubbles~\cite{GiblinHuiLimYang2010,KonstandinServant2011,GomesSimasNobregaAvelino2018,MilstedLiuPreskillVidal2022}, and non-perturbative  phenomena in quantum field theory~\cite{DashenHasslacherNeveu1974,DashenHasslacherNeveu1974a,Rajaraman1975,Evslin2019,Evslin2020}. 

Lots of kink solutions were constructed and studied in flat space-time~\cite{Vachaspati2007}, however, it is also possible to construct exact kink solutions in various gravity theories.  In the so-called thick brane models, for instance, our world is assumed to be a domain wall in a five-dimensional asymptotically anti de Sitter (AAdS) space-time~\cite{RubakovShaposhnikov1983,ArkaniHamedDimopoulosDvali1998,DeWolfeFreedmanGubserKarch2000,CsakiErlichHollowoodShirman2000,Gremm2000}. With the AAdS geometry, both gravity and matter fields can be localized on the wall without requiring compactification of the extra dimension~\cite{RandallSundrum1999,RandjbarDaemiShaposhnikov2000,KehagiasTamvakis2001,KoleyKar2005,LiuZhangZhangDuan2008}. 

Thick brane solutions were first found in minimally coupled Einstein-scalar systems, where field equations can be written as a group of first-order differential equations, after introducing the so-called superpotential~\cite{DeWolfeFreedmanGubserKarch2000,CsakiErlichHollowoodShirman2000,Gremm2000}. Usually, we have a freedom in choosing the form of the superpotential, thus, if the superpotential is properly chosen, exact thick brane solutions can be derived from the first-order equations~\cite{DeWolfeFreedmanGubserKarch2000,CsakiErlichHollowoodShirman2000,Gremm2000,BazeiaFurtadoGomes2004,BazeiaGomes2004,AfonsoBazeiaLosano2006}. Later, thick brane solutions were found in models with non-minimal couplings~\cite{AdamGrandiSanchezGuillenWereszczynski2008,BazeiaLosanoMenezes2008,LiuZhongYang2010,GuoLiuZhaoChen2012,LiuChenGuoZhou2012,ChenGuLiu2018,FuYuZhaoLiu2019}, with higher-order curvature terms~\cite{AfonsoBazeiaMenezesPetrov2007,LiuZhongZhaoLi2011,BazeiaLobaoMenezesPetrovSilva2014,BazeiaLobaoMenezes2015,ZhongLiuChenXie2014,ZhongLiu2016},  and in many other circumstances, see Refs.~\cite{DzhunushalievFolomeevMinamitsuji2010,Liu2018} for comprehensive reviews. 

Recently, it was found that thick brane like solutions also exist in some 2D gravity models. For example, in the following 2D dilaton gravity
\begin{equation}\label{action1}
	S=\frac{1}{\kappa}\int d^2x \sqrt{-g}\left(\varphi R-\frac{1}{2}\nabla^\mu\varphi\nabla_\mu\varphi+\kappa\mathcal{L}_m\right),
\end{equation}
static AAdS kink solutions can be constructed under the metric ansartz~\cite{Stoetzel1995,Zhong2021,ZhongLiLiu2021,Zhong2022a,FengZhong2022,Zhong2022,AndradeBazeiaLobaoJr.Menezes2022}:
\begin{equation}\label{metric}
	ds^2=-\text{e}^{2A(x)}dt^2+dx^2.
\end{equation}
Here $\kappa$ is the gravitational coupling constant, $\varphi$ is the dilaton field, $\mathcal{L}_m$ is the Lagrangian density of a scalar matter field $\phi$, which generates kink solutions, and $A(x)$ is the warp factor. The scalar matter field can be a canonical one or a noncanonical one, such as the so-called K-field~\cite{ZhongLiLiu2021,Zhong2022}.
In this model, the field equations have first-order formalism similar to those of the thick brane models in Refs.~\cite{DeWolfeFreedmanGubserKarch2000,CsakiErlichHollowoodShirman2000,Gremm2000}. Moreover, the linear perturbation equation for arbitrary static solutions of this model can be written as a Schr\"odinger-like equation with factorizable Hamiltonian, which ensures the stability of the solutions~\cite{Zhong2021,ZhongLiLiu2021,Zhong2022a}.

No doubt, the study of thick brane solutions extended our understanding about both gravity and kink.
However, having AAdS geometries  is just one possibility for gravitational kink solutions.  In principle, it is also possible for a gravitational kink to have other types of geometries. 

In a recent work~\cite{ZhongGuoLiu2024}, the authors found that by extending the dilaton kinetic term $\mathcal{X}\equiv-\frac{1}{2}\nabla^\mu\varphi\nabla_\mu\varphi$ into an arbitrary function $\mathcal{F}(\mathcal{X})$, one may obtain kink solutions with various geometries, depending on the form of $\mathcal{F}(\mathcal{X})$. For example, by taking $\mathcal{F}\propto \sqrt{-\mathcal{X}}$ the authors of Ref.~\cite{ZhongGuoLiu2024} found a kink solution with pure AdS$_2$ metric. More interestingly, the linear perturbation issue of this solution becomes a conformal quantum mechanics problem, if one of the model parameter takes a critical value. In this critical case, the linear perturbation equation is exactly solvable, and the corresponding quantum theory might be finite, as it has been discussed in the Liouville model, which has exactly the same perturbation equation~\cite{DHokerJackiw1982,DHokerFreedmanJackiw1983,DHokerJackiw1983}.

In this work, we explore another possibility of gravitational kinks, namely, kinks with asymptotically flat geometries. 
Such solutions may not be viable for braneworld consideration, but might be valuable for other purposes. For example, to numerically simulate the collision of gravitational kinks, one usually needs the metric to be asymptotically flat on at least one side, such that a smooth initial conditions can be constructed by superposing a kink and an antikink~\cite{TakamizuMaeda2006,TakamizuKudohMaeda2007,OmotaniSaffinLouko2011}.  Kink solutions with  asymptotically flat metrics on both sides may be relevant in the study of multi gravitating wall interactions, and may have different properties than those with AAdS or pure AdS metrics, but are seldom discussed so far.

For simplicity, we consider the case with two dimensions. The main idea is that we adopt the same 2D gravity model of Ref.~\cite{ZhongGuoLiu2024}, but chose another form of $\mathcal{F}(\mathcal{X})$.
Our model and solution will be given in the next section, after that we discuss the linear stability and scalar field localization issues of our solution. 

\section{The model and solution}\label{solutions}
The model of Ref.~\cite{ZhongGuoLiu2024} takes the following action
\begin{equation}\label{action2}
	S=\frac1\kappa\int d^2x\sqrt{-g}\left[\varphi R+\mathcal{F}(\mathcal{X})+\kappa\mathcal{L}_m\right].
\end{equation}
For simplicity, we assume that the scalar matter field to be a canonical one, namely, $\mathcal{L}_m=-\frac{1}{2}\nabla^\mu\phi\nabla_\mu\phi-V(\phi)$, where $V(\phi)$ is the interaction potential.

After variation, we obtain three field equations, namely, the scalar field equation
\begin{equation}\label{eom.scalar}
	\nabla_\lambda\nabla^\lambda\phi=V_{\phi},
\end{equation}
the dilaton equation
\begin{equation}\label{eom.dilaton}
	\nabla^\lambda\left(\mathcal{F_X}\nabla_\lambda\varphi \right)+R=0,
\end{equation}
and the Einstein equation
\begin{equation}
	\begin{aligned}\label{eom.metric}
		&\mathcal{F_X}\nabla_\mu\varphi\nabla_\nu\varphi-\frac12g_{\mu\nu}\left(-2\mathcal{F}+4\nabla_\lambda\nabla^\lambda\varphi\right)\\
		&+2\nabla_\mu\nabla_\nu\varphi+\kappa T_{\mu\nu}=0,
	\end{aligned}
\end{equation}
where $\mathcal{F_X}\equiv \frac{d \mathcal{F}}{d \mathcal{X}}$, and
\begin{equation}
	T_{\mu\nu}\equiv g_{\mu\nu}\mathcal{L}_m+\nabla_\mu\phi\nabla_\nu\phi,
\end{equation}
is the energy-momentum tensor. For the metric~\eqref{metric}, the dilaton equation~\eqref{eom.dilaton} becomes
\begin{equation}\label{dilaton.relation}
	\partial_x A=\frac{1}{2}\mathcal{F_X}\partial_x\varphi.
\end{equation}
Using this relation, the non-trivial components of the Einstein equation can be simplified as follows
\begin{eqnarray}\label{eom.00}
	-2\partial^2_x\varphi&=&\kappa(\partial_x\phi)^2,\\
	\label{eom.11}
	-2\partial^2_x\varphi+\mathcal{F}&=&\frac{1}{2}\kappa(\partial_x\phi)^2+\kappa V.
\end{eqnarray}
The scalar field equation
\begin{equation}
	\partial^2_x\phi+\partial_xA\partial_x\phi=\frac{dV}{d\phi},
\end{equation}
is not independent, and can be derived from Eqs.~\eqref{dilaton.relation}-\eqref{eom.11}. 

As shown in Ref.~\cite{ZhongGuoLiu2024}, Eqs.~\eqref{dilaton.relation}-\eqref{eom.11} have a simple first-order formalism:
\begin{eqnarray}\label{1st.phi}
	\partial_x\phi&=&\frac{dW}{d\phi},\\
	\label{1st.varphi}
	\partial_x\varphi&=&-\frac{\kappa}{2}W,\\
	\label{1st.A}
	\partial_x A&=&-\frac{\kappa}{4}\mathcal{F_X}W,\\
	\label{1st.V}
	V&=&\frac{\mathcal{F}}{\kappa}+\frac{1}{2}\left(\frac{dW}{d\phi}\right)^2,
\end{eqnarray}
where $W(\phi)$ is the superpotential function, whose form can be chosen arbitrarily. By taking appropriate $W(\phi)$ and $\mathcal{F(X)}$, exact kink can be derived form these first-order equations.

In this work we consider a model with 
\begin{equation}
	\mathcal{F(X)=X+\alpha X}^2,
\end{equation}
and
\begin{equation}
	W(\phi)=\sin {\phi}+c,
\end{equation}
where $\alpha$ and $c$ are two positive constant parameters. In this case, Eqs.~\eqref{1st.phi}-\eqref{1st.V} yield the following solution
\begin{eqnarray}\label{phi}
	\phi&=& \arcsin\left(\tanh ({x})\right),\\
	\label{varphi}
	\varphi&=&-\frac{c \kappa  x}{2}-\frac{1}{2} \kappa \ln (\cosh ({x})),\\
	\label{A}
	A&=&\frac{1}{32} \kappa  \Big\{2 c x \left[\alpha  \kappa ^2 \left(c^2+3\right)-4\right]\nonumber \\
	&+&2 \ln (\cosh({x})) \left[\alpha  \kappa ^2 \left(3 c^2+1 \right)-4\right]\nonumber \\
	&+&\alpha  \kappa ^2  \left[ \text{sech}^2({x})-6 c \tanh ( x)\right]\Big\},\\
	V&=&\frac{1}{64} \alpha  \kappa ^3 (c+\sin (\phi ))^4\nonumber\\
	&-&\frac{1}{8} \kappa  (c+\sin (\phi ))^2+\frac{1}{2}{\cos ^2(\phi )}.
\end{eqnarray}
Obviously, the parameter $\alpha$, hence the derivative coupling term of $\mathcal{F(X)}$, does not affect the solution of $\phi$ and $\varphi$, but it does affect the solution of the warped factor $A(x)$. 
To be more precise, the asymptotic behavior of  $A(x)$ is
\begin{equation}\label{asymptoticbehaviourofA}
	\lim_{x\to\pm\infty}A(x)=\frac{\kappa}{16}\left( B_{\pm}x+C_{\pm}\right),
\end{equation}
where 
\begin{equation}\label{Bpm}
	B_{\pm}=(c\pm1)\left[\alpha\kappa^2(c\pm1)^2-4 \right],
\end{equation}
and
\begin{equation}\label{Cpm}
	C_{\pm}=-\alpha\kappa^2 (c^2 \ln 8\pm3c+\ln 2)-\ln 16.
\end{equation}
Correspondingly, the scalar curvature at the boundary is
\begin{equation}
\label{curvarture}
	\lim_{x\to \pm\infty}R=-\frac{ \kappa ^2}{128}B_{\pm}^2.
\end{equation}

Obviously, when $c=0$ the warp factor is symmetric, and there exists a critical case where $\alpha=\alpha_{\text{c}}\equiv 4/\kappa^2$. If $\alpha\neq\alpha_{\text{c}}$, $B_{\pm}\neq 0$, the metric is AAdS$_2$, while if $\alpha=\alpha_{\text{c}}$, $B_{\pm}= 0$, the metric is asymptotically flat. 

\begin{figure*}
		{\includegraphics[width=16cm]{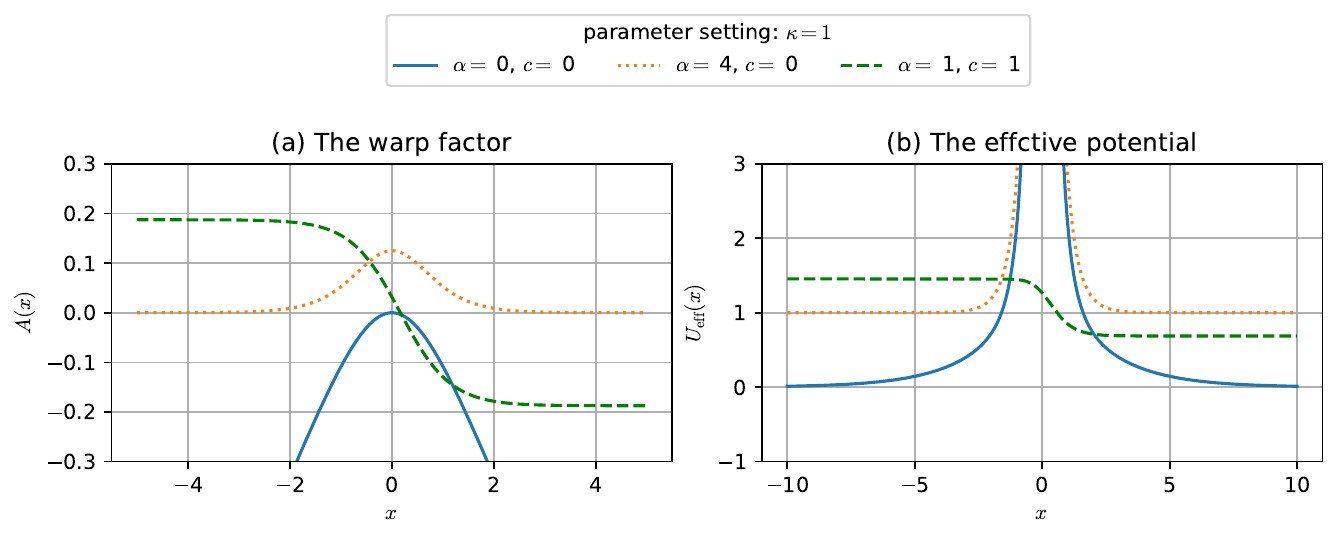}}
		\caption{Plots of (a) the warp factor $A(x)$, and (b) the effective potential $U_{\text{eff}}(x)$.}
		\label{FigAU}
\end{figure*}

On the other hand, if $c\ne0$ the warp factor is asymmetric. Especially, for $c=1$, $B_-=0$, thus the metric is already asymptotically flat on the left side.  In order the metric to be  asymptotically flat on the other side, we can simply take $\alpha=\tilde \alpha_c\equiv 1/\kappa^2$, such that $B_+=0$. In this case, $C_--C_+=6>0$,  the warp factor $A(x)$ has an antikink-like configuration, see Fig.~\ref{FigAU}~(a).

Now, let us turn to the stability issue.
\section{Linear perturbation}\label{perturbations}
Following Ref.~\cite{ZhongGuoLiu2024}, we consider the linear perturbation issue in the conformally flat coordinate
\begin{equation}\label{conformalflatmetric}
	ds^2=\text{e}^{2A(r)}(-dt^2+dr^2),
\end{equation}
where 
\begin{equation}\label{xrtransform}
	r(x)=\int_0^x d\tilde{x} \text{e}^{-A(\tilde{x})}.
\end{equation}
For simplicity, from now on, we use primes and over dots to denote the derivatives with respect to $r$ and $t$, respectively.

Suppose that $\{\bar{g}_{\mu\nu}(r), \bar\phi(r), \bar\varphi(r)\}$ constitute a  solution of the static field equations, we consider small perturbations around this solution:
\begin{equation}
	g_{\mu\nu}(t,r)=\bar{g}_{\mu\nu}(r)+\delta g_{\mu\nu}(t,r),
\end{equation}
\begin{equation}
	\phi(t,r)=\bar\phi(r)+\delta\phi(t,r),
\end{equation}
\begin{equation}
	\varphi(t,r)=\bar{\varphi}(r)+\delta\varphi(t,r).
\end{equation}
To start with, we rewrite metric perturbation as~\cite{Zhong2021,ZhongGuoLiu2024}
\begin{equation}
	\delta g_{\mu\nu}=\text{e}^{2A}h_{\mu\nu}=\text{e}^{2A}\left[\begin{matrix}
		h_{00} & \Phi\\
		\Phi & h_{rr}
	\end{matrix}\right]
\end{equation}
such that 
\begin{equation}
	\delta g^{\mu\nu}=-\text{e}^{2A}h^{\mu\nu}=\text{e}^{2A}\left[\begin{matrix}
		h_{00} & -\Phi\\
		-\Phi & h_{rr}
	\end{matrix}\right],
\end{equation}
where the  indices of $h_{\mu\nu}$ are raised by $\eta^{\mu\nu}$.

After linearizing the field equations, one obtains three independent perturbation equations. Two of them come from the linearized Einstein equation, namely, the (1, 1) component  
\begin{equation}\label{perturb11}
	\Xi=\kappa\frac{\phi'}{\varphi'}\left[\delta\phi'+\delta\phi\left(\frac{\varphi''}{\varphi'}-\frac{\phi''}{\phi'}-\mathcal{F_{XX}X}\varphi'\right)\right],
\end{equation}
and the (0, 1) component
\begin{equation}\label{perturb01}
	h_{rr}=\kappa\frac{\phi'}{\varphi'}\delta \phi.
\end{equation} Here we have defined a new variable $\Xi=2\dot\Phi-h_{00}'$ and taken the dilaton gauge $\delta\varphi=0$ to eliminate the gauge degrees of freedom. 
The last equation comes from the linearized scalar field equation
\begin{eqnarray}\label{perturbEoM}
	&&-\ddot{\delta \phi }+\delta \phi ''-\frac{  \phi'''}{\phi '}\delta \phi+\frac{\mathcal{F}_{\mathcal{X}}\varphi '  \phi ''}{\phi '}\delta \phi \nonumber\\
	&&-\frac{\phi ' }{2}h_{rr}'- \phi ''{h_{rr}}+\frac{\phi '}{2}\Xi  =0.
\end{eqnarray}
After eliminate $\Xi$ and $h_{rr}$ by using Eqs.~\eqref{perturb11} and~\eqref{perturb01}, Eq.~\eqref{perturbEoM} becomes~\cite{ZhongGuoLiu2024}:
\begin{equation}
	\ddot{\delta\phi}-\delta\phi''+U_{\text{eff}}\delta\phi=0,
\end{equation}
where the effective potential is
\begin{equation}
	U_{\text{eff}}=\frac{f''}{f}, \quad f= \frac{\phi'}{\varphi'}.
\end{equation}

The time-independence of $U_{\text{eff}}$ allows us to use mode expansion, $\delta\phi=\sum_{n} \Theta_n(r)\text{e}^{i\omega_n t}$, which leads to a Schr{\"o}dinger-like equation of $\Theta_n(r)$ 
\begin{equation}\label{schrodingerequation}
	\hat{H}\Theta_n=\omega_n^2\Theta_n,
\end{equation}
where the Hamiltonian operator is
\begin{equation}
	\hat{H}=-\frac{d^2}{dr^2}+U_{\text{eff}}.
\end{equation}
Note that the Hamiltonian operator can be factorized as a product of two hermitian conjugate operators:
\begin{equation}
	\hat{H}=\hat{A}\hat{A}^\dagger,
\end{equation}
where
\begin{equation}
	\hat{A}=\frac{d}{dr}+\frac{f'}{f},\quad \hat{A}^\dagger=-\frac{d}{dr}+\frac{f'}{f}.
\end{equation}
The zero mode $\Theta_0$, i.e., the one with eigenvalue $\omega_0=0$, satisfies $\hat{A}^\dagger \Theta_0=0$ and takes the following form:
\begin{equation}\label{zeromode}
	\Theta_0\propto f.
\end{equation}

Since there is no analytical expression of transformation~\eqref{xrtransform} in general, it is useful to transformation $U_{\text{eff}}$ back to $x$-coordinate
\begin{equation}
	U_{\text{eff}}(x)=\text{e}^{2A}\left(\partial_xA\frac{\partial_xf}{f}+\frac{\partial_x^2f}{f}\right),
\end{equation}
where $f(x)=\partial_x\phi/\partial_x\varphi$. Substituting Eqs.~\eqref{phi}-\eqref{A} into above equation, one obtains the expression of $U_{\text{eff}}(x)$ immediately. 

For the kink solution with $c=0$, $\alpha=\alpha_{c}=4/\kappa^2$, the effective potential reads
\begin{eqnarray}
	U_{\text{eff}}(x)&=&2 \left[-\kappa +(\kappa +8) \cosh (2 x)+\cosh (4 x)+7\right] \nonumber\\
	&\times & \text{e}^{\frac{1}{4} \kappa  \text{sech}^2(x)+4 x}\left(1-\text{e}^{4 x}\right)^{-2},
\end{eqnarray}
which has a singularity at the origin $x=0$, and $\lim_{|x|\to\infty}U_{\text{eff}}(x)=1$. The spectrum is a continuous interval $(1,+\infty)$.

While for the solution with $c=1$, $\alpha=\tilde{\alpha}_c=1/\kappa^2$, the effective potential becomes
\begin{eqnarray}
	U_{\text{eff}}(x)&=&\frac{1}{16}  (3 \kappa +\kappa  \tanh (x)+8 \cosh (2 x)+8) \nonumber\\
	&\times&\text{e}^{\frac{1}{16} \kappa  \left(\text{sech}^2(x)-6 \tanh (x)\right)}\text{sech}^2(x),
\end{eqnarray} 
which has no singularity. In this case, the effective potential is positive-definite and its asymptotical behaviour is $\lim_{x\to\pm\infty}U_{\text{eff}}(x)=\exp({\pm{3}\kappa/8})$. Hence the spectrum is a continuous interval $(\text{e}^{-3\kappa/8},+\infty)$. 

Figures of $U_{\text{eff}}(x)$ can be found in Fig.~\ref{FigAU}~(b). Just as the AAdS kink solution, both of the  asymptotically flat kinks have no normalizable zero mode or other bounded oscillation modes in the linear spectra.
In fact,  the normalized condition for  the zero mode is 
\begin{eqnarray}
	&&\int_{-\infty}^{+\infty}dr|f(r)|^2\nonumber\\
	&=&\int_{-\infty}^{+\infty}dx\text{e}^{-A}\left(\frac{2}{\kappa}\frac{\text{sech}(x)}{\tanh(x)+c}\right)^2<\infty,
\end{eqnarray}
which cannot be satisfied if the metric is asymptotically flat, because the integrand diverges either at $x=0$ if $c=0$, $\alpha=\alpha_c$, or at $x=-\infty$, if $c=1$, $\alpha=\tilde\alpha_c$. 

\section{Localization of scalar field}\label{localization}
As in the case of thick brane models, it is interesting to consider the propagation of bulk matter on backgrounds of kink solutions, and see if the matter fields are trapped around the kink~\cite{LiuGuoFuLi2011,BajcGabadadze2000}. For simplicity, we only consider the propagation of scalar fields in this work.

If one starts with a massless scalar field
\begin{equation}\label{actionmassless}
	S_{\text{scalar}}=\int d^2x\sqrt{-g}\left\{-\nabla_\mu\Phi\nabla^\mu\Phi\right\},
\end{equation}
then the equation of motion
\begin{equation}\label{eomofscalar}
	\frac{1}{\sqrt{-g}}\partial_{\mu}\left(\sqrt{-g}g^{\mu\nu}\partial_{\nu}\Phi\right)=0
\end{equation}
becomes
\begin{equation}
	\label{EqMassless}
	(-\partial_t^2+\partial_r^2)\Phi=0
\end{equation} 
in the conformally flat coordinate.
Equation~\eqref{EqMassless} is just the equation for a free massless scalar field in flat space-time, and therefore, has no trapped modes around the kink. This is because in 2D space-time, the action of massless minimally coupled scalar field is conformally invariant~\cite{Jacobson2003}. 
Thus, to trap scalar matters around 2D gravitating kinks, we need some other mechanisms. 

\subsection{Mechanism I}
One mechanism for trapping scalar matter is to introduce a scalar-kink interaction of the following form~\cite{Vachaspati2007}:
\begin{equation}\label{interaction}
	\mathcal{L}_{\text{int}}=-\frac{\lambda}{2}\phi^2\Phi^2,
\end{equation}
with which Eq.~\eqref{EqMassless} becomes
\begin{equation}
	(\partial_t^2-\partial_r^2+{\lambda}\text{e}^{2A}\phi^2)\Phi=0.
\end{equation}
After conducting the mode expansion $\Phi=\sum_{n}{\psi}_n(r)\text{e}^{im_nt}$, the spatial components ${\psi}_n$ satisfy a Schr\"{o}dinger-like equation
\begin{equation}
	\label{equationofsk}
	(-\partial_r^2+V_{\text{eff}}){\psi}_n=m_n^2{\psi}_n,
\end{equation}
where the effective potential is
\begin{equation}
	V_{\text{eff}}(r)=\lambda \text{e}^{2A(r)}\phi^2(r).
\end{equation} 
We solve Eq.~\eqref{equationofsk} numerically after inserting our solutions obtained in Sec.~\ref{solutions}, and find that both of the two asymptotically flat kink solutions support two bound states. For the solution with $\alpha=4$ and $c=0$, the eigenvalues of the bound states are $m^2_{1}\approx0.901$ and $m^2_{2}\approx2.089$. While, for the one with $\alpha=c=1$, the discrete eigenvalues are $m^2_{1}\approx0.791$ and $m^2_{2}\approx1.655$. The figures of the effective potentials and the wave functions of the bound states are drawn in the upper panel of Fig.~\ref{FigSK}. 

In our calculation, we have set $\kappa=\lambda=1$, and there is no localized zero mode in the spectra. In fact, the effective potential $V_{\text{eff}}\geq0$ for $\lambda>0$, thus, the absence of zero mode is general for mechanism I with positive $\lambda$. 

\subsection{Mechanism II}
\begin{figure*}
		{\includegraphics[width=16cm]{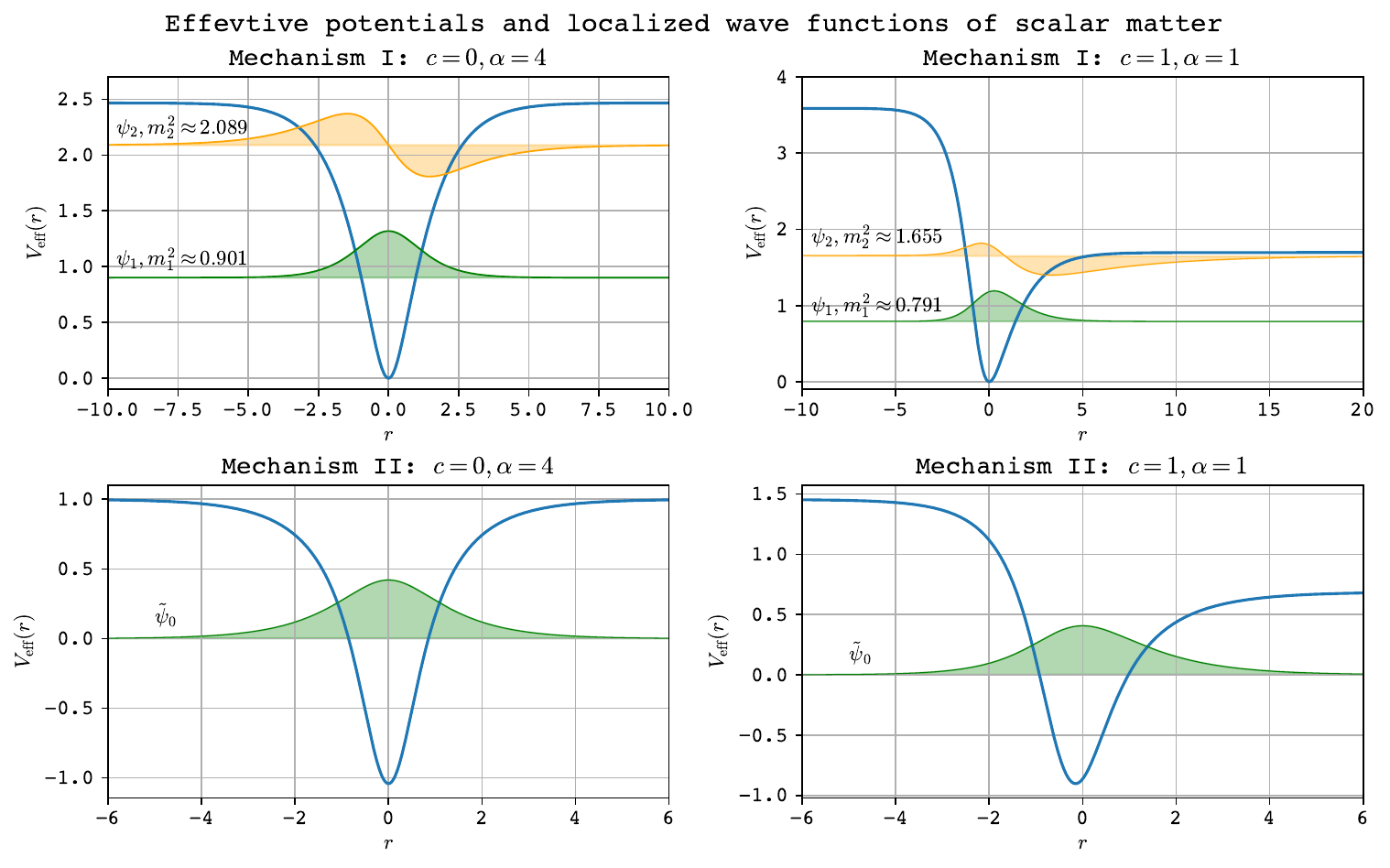}}
		\caption{\label{FigSK} Effective potentials and localized wave functions of scalar matter. We have set $\kappa=\lambda=1$. Note that the wave functions corresponding to the bound states are vertically shifted with the relevant eigenvalues.}
\end{figure*}

Alternatively, one may start by assuming that the scalar  field has a noncanonical kinetic term~\cite{AraiBlaschkeEtoSakai2019}: 
\begin{equation}
	\label{JRaction}
	S_{\text{scalar}'}=\int d^2x\sqrt{-g}\left\{-\beta(\phi)^2\nabla_\mu\Phi\nabla^\mu\Phi\right\},
\end{equation}
where  $\beta(\phi)$ is a function of the kink field $\phi$.  This mechanics is called Jackiw-Rebbi-like mechanism~\cite{AraiBlaschkeEtoSakai2019}, which extends the Jackiw-Rebbi mechanism~\cite{JackiwRebbi1976}. For action \eqref{JRaction}, the equation of motion reads
\begin{equation}
	\frac{1}{\sqrt{-g}}\partial_{\mu}\left(\sqrt{-g}g^{\mu\nu}\beta(\phi)^2\partial_{\nu}\Phi\right)=0.
\end{equation}
After conducting the mode expansion $\Phi=\sum_n {\psi}_n(r)\text{e}^{im_nt}$, one finds that the spatial components ${\psi}_n$ satisfy
\begin{equation}\label{JRmechanism0}
	\beta\partial^2_r{\psi}_n-2\partial_r\beta\partial_r{\psi}_n=m^2_n{\psi}_n.
\end{equation}
If we define
\begin{equation}
	{\psi}_n\equiv \tilde{\psi}_n {\beta},
\end{equation}
then Eq.~\eqref{JRmechanism0} becomes a  Schr\"odinger-like equation~\cite{AraiBlaschkeEtoSakai2019}:
\begin{equation}
\label{JRmechanism}
	(-\partial_r^2+V_{\text{eff}})\tilde{\psi}_n=m_n^2\tilde{\psi}_n,\quad V_{\text{eff}}=\frac{\partial_r^2\beta}{\beta}.
\end{equation}
Similar to Eq.~\eqref{schrodingerequation}, Eq.~\eqref{JRmechanism} has a zero mode $\tilde{\psi}_0\propto\beta$. Thus, if $\beta(\phi(r))$ is a square integrable function of $r$, the scalar $\Phi$ has a normalizable zero mode. 

As an example, we take 
\begin{equation}\label{coupling}
	\beta(\phi)=\phi^2-\frac{\pi^2}{4},
\end{equation}
Obviously, as $x\to\pm\infty$, $\beta(\phi)\to0$. 
The normalization condition of the zero mode is
\begin{equation}
\int_{-\infty}^{+\infty}dr \tilde{\psi}_0^2=	\mathcal{N}^{2}\int^{+\infty}_{-\infty}dx\text{e}^{-A}\beta^2=1,
\end{equation}
where  $\mathcal{N}$ is the normalization constant. For the asymptotically flat kink solution with $\alpha=4$ and $c=0$, we get $\mathcal{N}\approx0.284$. While for the one with $\alpha=c=1$, the normalization constant $\mathcal{N}\approx0.275$. These results are obtained by taking $\kappa=1$. In this case, our numerical calculation shows that the zero mode is the only bound state in the spectra, see the bottom panel of Fig.~\ref{FigSK}.

\section{Conclusion}\label{conclusion}
In this work, we studied a generalized 2D dilaton gravity model, in which the dilaton field has a noncanonical kinetic term $\mathcal{X}+\alpha \mathcal{X}^2$. We found that for some special parameters, there exist analytical static kink solutions with asymptotically flat metrics, which to our knowledge, were not reported before. The linear perturbation analysis indicates that our solutions have positive continuous linear spectra, and therefore, are stable against small linear perturbations.

We also studied the propagation of bulk scalar matter fields on our gravitating kink backgrounds. We find that a minimally coupled massless scalar field propagates freely on the kink background, no matter what the metric is. This is different from the case of 5D thick branes, where a minimally coupled bulk scalar field usually feels an effective potential around the kink, thanks to the AAdS geometry. The reason behind is that in 2D space-time, the action of a minimally coupled massless scalar field is conformally invariant.

In order to trap at least a few modes of the scalar field $\Phi$, we considered two different mechanisms for scalar matter localization. The first mechanism assumes a scalar-kink interaction term $\frac{\lambda}{2}\phi^2\Phi^2$ on top of the canonical kinetic term of $\Phi$. The other is the so-called Jackiw-Rebbi-like mechanism, which assums a noncanonical kinetic term of $\Phi$. After inserting our solutions and taking $\kappa=\lambda=1$, we found that the first mechanism allows two localized modes with positive eigenvalues, but has no zero mode. With the second mechanism, we have a localized zero mode, which is the only bound state in the spectra. 

It would be interesting to study the collision of gravitating kinks with asymptotically flat metrics, or to study the localization of matter fields with other spins. We leave these questions to our future works.

\begin{acknowledgments}
This work was supported by the National Natural Science Foundation of China (Grant No. 12175169).
\end{acknowledgments}

%
%


\end{document}